# Pressure-induced superconductivity in Iron pnictide compound SrFe$_2$As$_2$

Kazumi IGAWA, Hironari OKADA, Hiroki TAKAHASHI, Satoru MATSUISHI[1], Yoichi KAMIHARA[2], Masahiro HIRANO[1,2], Hideo HOSONO[1,2], Kazuyuki MATSUBAYASHI[3], and Yoshiya UWATOKO[3]

*Department of Physics, College of Humanities and Sciences, Nihon University, Sakurajosui, Setagaya-ku, Tokyo 156-8550*

*1 Frontier Research Center, Tokyo Institute of Technology, 4259 Nagatsuda, Midori-ku, Yokohama 226-8503*

*2 ERATO-SORT, JST, in Frontier Research Center, Tokyo Institute of Technology, 4259 Nagatsuda, Midori-ku, Yokohama 226-8503*

*3 The Institute for Solid State Physics, Tokyo University, Kashiwanoha, Kashiwa, Chiba 277-8581*

The discovery of the Fe-based superconductors had a significant impact in the field of condensed matter physics.[1,2] The atomic substitution and pressure effect have promoted a significance increase of $T_c$.[3-6] Recently, $A$Fe$_2$As$_2$ ($A$=Ca, Sr and Ba) with the tetragonal ThCr$_2$Si$_2$ structure have been discovered to show superconductivity through introducing additional charge carriers by the chemical substitution. For example, substitution K for Ba in BaFe$_2$As$_2$ induces superconductivity at 38 K.[7] The undoped $A$Fe$_2$As$_2$ ($A$=Ca, Sr and Ba) compounds exhibit a tetragonal to orthorhombic transition below room temperature and magnetic ordering (SDW) in the orthorhombic phase,[8-10] such as LaFeAsO. The chemical substitution introduces the charge carriers and suppresses the structural and magnetic transitions.[8] Among the undoped compounds of this family, the smallest volume member CaFe$_2$As$_2$ was reported to exhibit the pressure-induced superconductivity with $T_c$ = 13 K above 0.5 GPa.[11,12] The superconductivity appeared after the suppression of the structural and magnetic transition, and it disappeared above 0.9 GPa. For CaFe2As2 a tetragonal to tetragonal transition was also reported. Above 0.3 GPa, the large decrease of electrical resistivity was detected,[13,14] which was recognized as a "collapsed" tetragonal phase by performing a high-pressure neutron diffraction measurement.[13] The pressure-induced superconductivity has also been reported for SrFe$_2$As$_2$ and BaFe$_2$As$_2$ by measuring a magnetic susceptibility using a diamond anvil cell.[15] The peak values of $T_c$ were 29 K at $P$ = 3.5GPa and 27 K at $P$ =3.0 GPa, for BaFe$_2$As$_2$ and SrFe$_2$As$_2$, respectively. The electrical resistivity measurements were also performed for these compounds, where the suppression of both structural and magnetic transitions and the resistivity loss associated with superconductivity were observed under high pressure.[11-14,16] However, no zero resistivity were detected in their measurements. In this study, the high-pressure resistivity measurements for SrFe$_2$As$_2$ up to 14 GPa have been carried out. The suppression of the structural and magnetic transitions under high pressure and the pressure-induced superconductivity are presented.

Polycrystalline SrFe$_2$As$_2$ was prepared by a solid-state reaction reported elesewhere.[17] Electrical resistivity measurements under high pressure were performed by means of a standard dc four-probe method. The hydrostatic pressures up to 1.5 GPa and 8 GPa were applied by a piston-cylinder device and a cubic anvil cell,[18] respectively, using Fluorinert (FC-77:FC-70 = 1:1) as a pressure transmitting medium. A diamond anvil cell (DAC) made of CuBe alloy was used for electrical resistivity measurements at pressures up to 14 GPa using powdered NaCl as the pressure-transmitting medium.

Figure 1 shows the $\rho$-$T$ curve at each pressure using the cubic anvil cell. The rapid decrease in the electrical resistivity was observed at around 200 K for the data at 1.5 GPa. This decrease is considered to be caused by the tetragonal to orthorhombic transition and is observed at the almost same temperature as reported value of 205 K.[9,17] The magnetic transition was reported just below the structural transition temperature. In this paper, the characteristic temperature $T_0$ defined at the peak value of d$\rho$/d$T$ is adopted to stand for these transitions, which is indicated by arrows in Figure 1. The $T_0$ decreases with a rate of -10 K/GPa, and could not be detected above 8 GPa. This decreasing rate is rather smaller than the reported value of -13 K/GPa.[17] On the other hand, the electrical resistivity decreases steeply at around 25 K at 8 GPa and down to a zero value at around 10 K. Moreever, the enhancement of resistivity was observed with increasing the electrical current from 0.1 mA to 1.0 mA below 25 K. This seems to be attributed to the destruction of superconductivity due to the electrical current. Therefore, this decrease of resistivity is concluded as the superconducting transition. Although a steep decrease in resistivity is still detectable down to 1.5 GPa below 8 GPa, there is no sign of the zero resistivity. However, we consider these anomalies in $\rho$-$T$ curves as superconducting transitions, because these anomalies are observed as a similar fashion for each pressures. In this way, the superconducting transition temperatures ($T_c$) of SrFe$_2$As$_2$ are plotted in Figure 2, together with the $T_0$. The $T_c$ is defined at the onset of the transition, determined as shown in the inset of Figure 1. The $T_c$ increases steeply up to 38 K below 2 GPa, and decreases slowly with further compression. The shape of $T_c$ ($P$) is similar to the previously reported phase diagram of SrFe$_2$As$_2$[15] determined by the magnetic susceptibility measurements. However, the $T_c$ is higher and the pressure range of the appearance of superconductivity is wider in this measurement. Since the electrical resistivity measurement is often more sensitive for detecting the superconductivity than the magnetic susceptibility measurement, we consider the quantitative difference of $T_c$ -$P$ curve is due to the method of determination of $T_c$. Gooch et al.[22] examined K$_{1-x}$Sr$_x$Fe$_2$As$_2$ system under high pressure and reported that the pressure dependence of $T_c$ is similar to the cuprate superconductors. That is, the pressure effect on $T_c$ for the underdoped K$_{1-x}$Sr$_x$Fe$_2$As$_2$ shows positive pressure coefficient, while for the overdoped ones it shows the negative pressure coefficient. In this study, the decrease of $T_c$ with increasing pressure may be due to the overdoping. It is noteworthy that if the $T_0$ is related to the SDW transition, the superconductivity of SrFe$_2$As$_2$ may coexist with the SDW phase, as pointed out in the previous reports.[19] Thus, it is possible that the positive pressure effect on $T_c$ is caused by the suppression of the magnetic phase and the increase of charge carriers, while the negative pressure effect in the higher pressure region is caused by the carrier doping for the overdoping state. The experiments for the electronic state under high pressures are urgently needed.

**Figure captions**

Fig.1. (Color online) The $\rho$-$T$ curves of SrFe$_2$As$_2$ under high pressure up to 8 GPa, using the cubic anvil cell. Arrows indicating the characteristic temperature $T_0$ shift toward the lower temperature by applying pressure. The inset shows enlargement in the low temperature range.

Fig.2. (Color online) The $P$-$T$ phase diagram of SrFe$_2$As$_2$ obtained from electrical resistivity measurements. Run 1 (close triangle) show the results using the piston-cylinder device. Run 2 (close diamond) and Run 3 (close square) show the run using the DAC and the cubic anvil cell, respectively. The broken-line curve is guide for the eye.

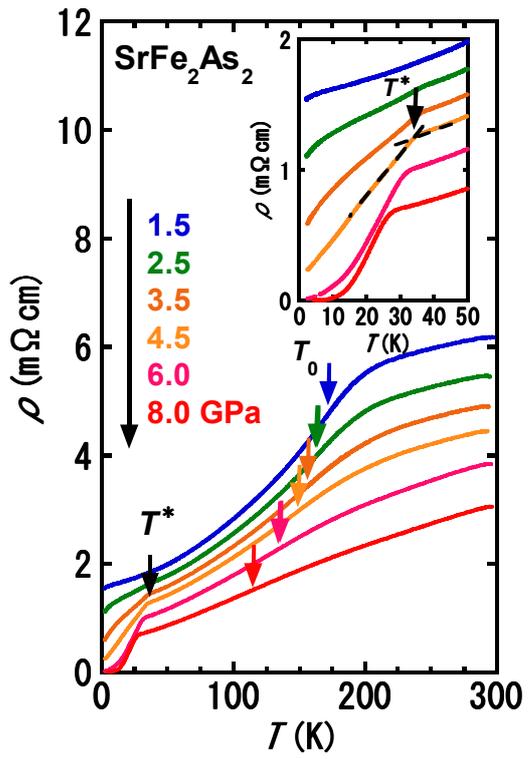

Fig. 1

K. Igawa *et al.*

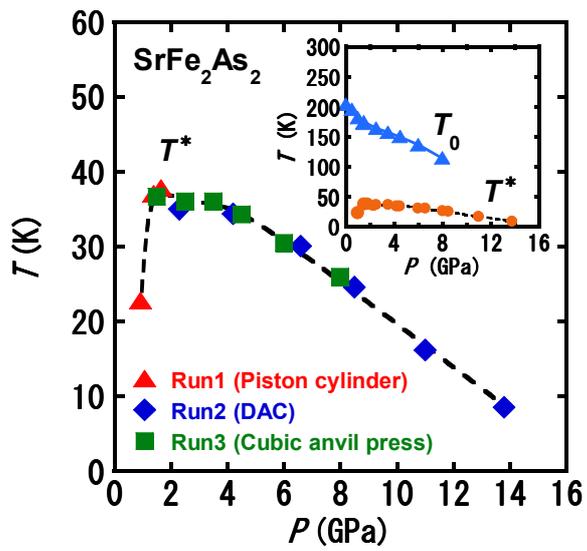

Fig. 2

K. Igawa *et al.*